\documentclass[conference]{IEEEtran}

\IEEEoverridecommandlockouts
\usepackage{graphicx}
\usepackage{enumitem}
\newtheorem{hypothesis}{Hypothesis}

\title{\LARGE \bf The Myth of Meritocracy and the Matilda Effect in STEM:\\ Paper Acceptance and Paper Citation }

\author{Joana Fonseca 
\thanks{Joana Fonseca is with the ACCESS Linnaeus Centre, School of Electrical Engineering and Computer Science. KTH Royal Institute of Technology, SE-100 44 Stockholm, Sweden. {\tt\small jfgf@kth.se}.
} }

\begin{document}
\maketitle
\thispagestyle{empty}
\pagestyle{empty}

\begin{abstract}

Biases against women in the workplace have been documented in various studies. There is also a growing body of literature on biases within academia.
But particularly in STEM, due to the heavily male-dominated field, studies suggest that if one's gender is identifiable, women are more likely to get their papers rejected and not cited as often as men.  

We propose two simple modifications to tackle gender bias in STEM that can be applied to (but not only) IEEE conferences and journals. Regarding paper acceptance, we propose a double-blind review, and regarding paper citation, we propose one single letter to identify the authors' first names, followed by their family names. We also propose other modifications regarding gender bias in STEM and academia and encourage further reforms supported by current research on this topic with gender-segregated data.

\end{abstract}

\section{INTRODUCTION}

While most of us agree that there are systemic inequalities related to gender, the magnitude of these inequalities is rarely clear.  
We know that there is a gender pay gap, but not many of us, however, realize to what extent gender bias has a profound, negative impact on women’s career success. Perhaps because we are so used to the way things are and have always been? \cite{perezreview}.

To answer this question, we provide some background knowledge on gender inequality in the workplace and career, focusing on academia and, particularly, STEM as a male-dominated field. We also explain the Matilda effect (first introduced by Rossiter in 1993 on \cite{matildafirst}) and its negative consequences on women's career success.


The myth of meritocracy drives gender inequalities at work – those with ‘merit’ get hired, promoted, and get a higher salary, which means that there are fewer women on top and that women’s salaries are lower. Even in countries that are leading the way in this (e.g., Scandinavian countries), hiring and promotion are still impacted by gender bias, for example, in academia \cite{scandinavia}. 

Perez wrote a chapter called \textit{The Myth of Meritocracy} in which she divides our failure to understand and acknowledge merit in the workplace into fourth main problems. \cite{perez2019}. They are the lack of sex-disaggregated data, the belief in merit, the rise of new technologies that increase the gender gap, and the lack of evidence-based solutions. In the current paper, we will use diverse research papers that collected and analyzed sex-disaggregated data on women's careers and paper acceptance and citation. These papers disprove many meritocracy beliefs, and with their results, I propose two evidence-based solutions.

\bigskip

Although women fill nearly half of all jobs in the U.S. economy, they hold less than 25\% of STEM jobs. This has been the case throughout the past decade, even as college-educated women have increased their share of the overall workforce \cite{stem}.
In 2017 UNESCO commissioned a report on women's and girls' education in STEM \cite{unesco2017}. Here they state that girls’ under-representation in science, technology, engineering, and mathematics (STEM) education is deep-rooted and puts a detrimental brake on progress toward sustainable development. The focus is on the need to understand the drivers behind this situation to reverse these trends.

\bigskip

This gender bias is so persistent it has been acknowledged and named the \textit{Matilda effect}.
The Matilda effect is a bias against women in male-dominated fields that systematically unrecognises them \cite{mathilda}. Specifically among scholars,
this negative evaluation bias should subsequently reduce interest in connecting or working with the evaluated individual. Male researchers exhibited a bias toward citing same-sex authors more frequently, whereas female researchers cited authors of both sexes in proportion to the pool of publications available for being referenced. 

However, other work has shown that both sexes tend to evaluate women less favorably in performance contexts and that women may prefer a male boss over a female boss even more strongly than men \cite{Lever2011}. The Matilda effect has been found in various areas of academia, including in the 2011 European Commission Meta-analysis of Gender and Science Research \cite{eu2012}.

In the following chapter, we'll provide data-segregated results that support the hypothesis of the Matilda effect in academia and STEM.

\section{RESULTS}\label{s: main}
 
Based on the considerations analyzed in the Introduction, we propose the following three hypotheses. The first one concerns the existence of gender inequality in contributions in academia, and the two last ones consider the efficacy of solutions to tackle the problem the first hypothesis presents.

\begin{hypothesis}[Matilda effect on academia and STEM]\label{h1}
    Male authors’ scholarly contributions are associated with greater scientific quality than female authors’ scholarly contributions.
\end{hypothesis}

\begin{hypothesis}[Double blind review and the Matilda effect]\label{h2}
    Implementing a double-blind review for paper acceptance in conferences and journals significantly increases paper acceptance for women-first authors while decreasing paper acceptance for men-first authors.
\end{hypothesis}

\begin{hypothesis}[First letter of first name and the citation gap]\label{h3}
    Implementing the first letter only for the authors' first name in a contribution significantly increases paper citation for women first authors.
\end{hypothesis}

In Subsection \ref{PA}, we review two papers that evaluate paper acceptance before and after the implementation of double-blind review. These should provide relevant arguments towards Hypothesis \ref{h1} and \ref{h2}.

In Subsection \ref{PC}, we review two papers that evaluate paper citations with and without identifiable gender. These should provide relevant arguments towards Hypothesis \ref{h1} and \ref{h3}.

In Subsection \ref{EG}, we review one paper that evaluates exam grading with and without identifiable gender. This topic does not refer to paper acceptance or citation. However, it's still relevant for us as teachers and examiners, and it might as well give some parallel conclusions toward paper acceptance. So, to some extent, this should provide relevant arguments towards Hypothesis \ref{h1} and \ref{h2}.

Note that counterarguments to these hypotheses will be discussed in Section \ref{discussion}.
 

\bigskip
\subsection{Paper acceptance: before and after implementing double-blind review} \label{PA}
 
\medskip
\section*{Double-blind review favors increased representation of female authors}
\medskip 
 
In 2007, Budden et al. \cite{BUDDEN20084} wrote a paper that analyses the percentage of women first-authored publications in the journal Behavioral Ecology (\textit{BE}) before and after the implementation of double-blind review in 2001. 

\begin{figure}
\centering
\includegraphics[width=.46\textwidth,clip]{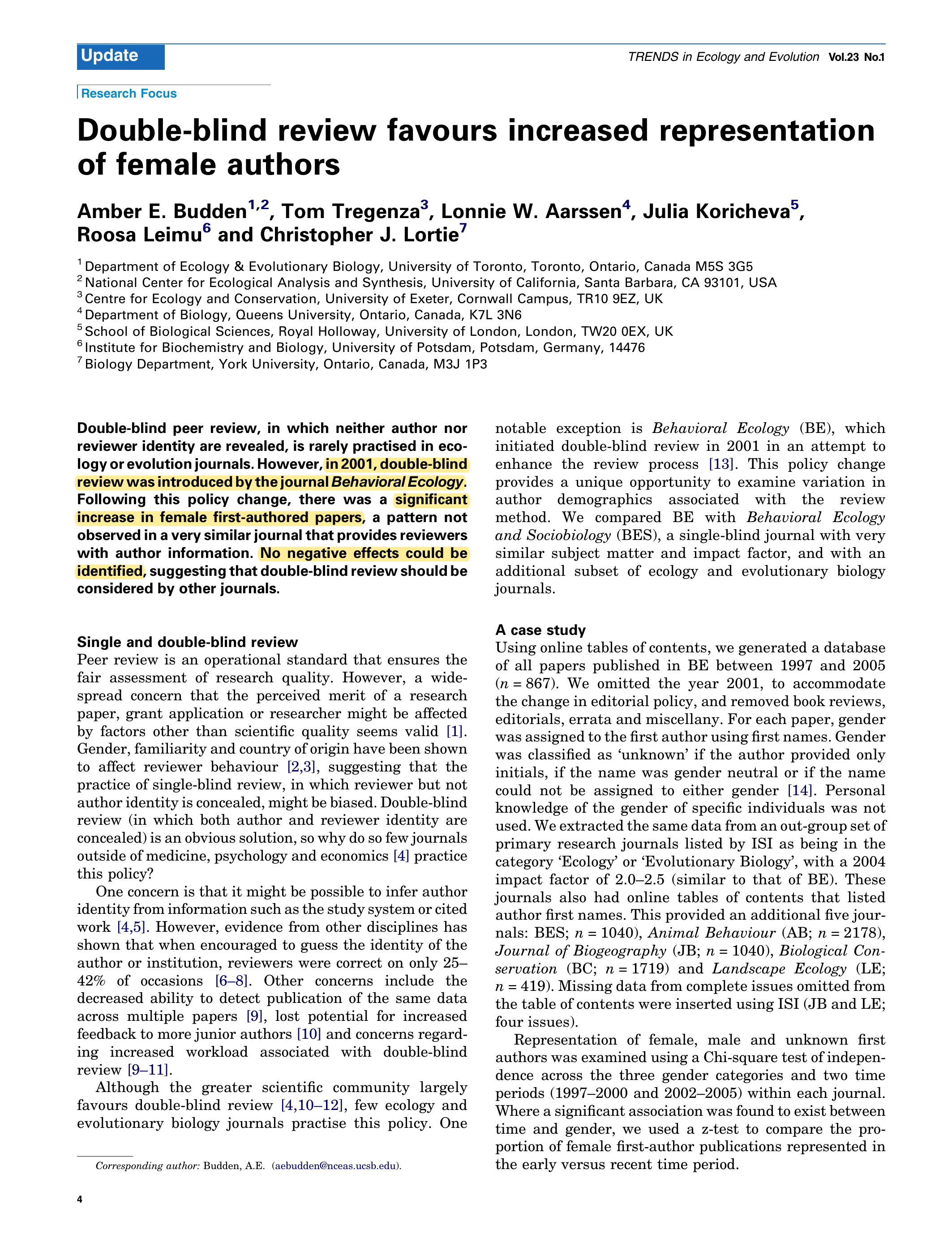}
\caption{Papers published in Behavioral Ecology by first-author gender. (a) Total number of papers published in BE in the four years before and after the implementation of a double-blind review policy in 2001. (b) Percentage change in author representation.} \label{BE}
\end{figure}

The data used was a database of all papers published in \textit{BE} and \textit{BES} (according to the authors, this is a very similar journal, attracting the same type of researchers, but that did not implement DBR) between 1997 and 2005 ($n = 867$).
Representation of female, male, and unknown first authors was examined across two time periods: 1997–2000 (before DBR) and 2002–2005 (after DBR) within each journal, as seen in Fig. \ref{BE}.

Note that in the four years following the introduction of
double-blind review, \textit{BE} published more papers by both genders. However, the magnitude of this difference was significantly larger for females than for males. Following
a double-blind review, there was a 7.9\% increase in the
proportion of papers with a female first author and a corresponding decrease in papers with a male first author. 

If the proportion of women in the field has increased or increased in productivity, we would predict a commensurate
change in authors publishing in \textit{BES} (the other journal which doesn't have DBR). However, no significant difference in gender representation was found across the same time period in \textit{BES}, which strongly suggests that the change is directly related to review policy.

Note that this difference of 7.9\% in the proportion of female first-authored papers following the implementation of double-blind review in \textit{BE} is three times greater than the recorded increase in female ecology graduates in the USA across the same time period and represents a 33\% increase in the representation of female authors.

\medskip
\section*{Double-blind reviewing at EvoLang 11 reveals
gender bias}
\medskip

More recently, in 2016, Roberts et al. \cite{RobertsVerhoef} analyzed the impact of introducing a double-blind review in EvoLang 11 (Evolution of Language conference) by comparing it to EvoLang 9 and 10. Main effects were found for first author gender by conference.

The authors segregated the data into women and men to evaluate the \textit{Matilda effect} (recall: bias against women in male-dominated fields) and into student and non-student to evaluate the \textit{Matthew effect} (recall: bias favoring well-established academics).

\begin{figure}
\centering
\includegraphics[width=.46\textwidth,clip]{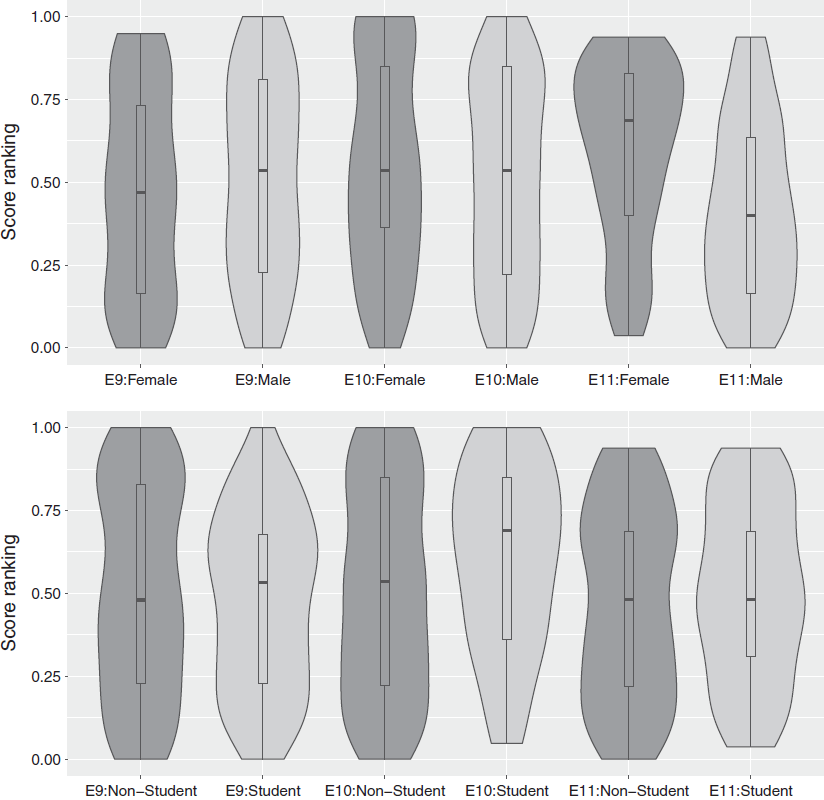}
\caption{Upper: Differences in ranking by gender of the first author. Lower: Differences in ranking by student status of the first author.} \label{Evolang}
\end{figure}

The results in Fig. \ref{Evolang} showed that there was little difference in ranks for papers with male or female first authors for
EvoLang 9 (difference in $means=0.04$, $t=0.87$, $P=0.386$) or EvoLang 10 (difference in $means=0.04$, $t=0.75$, $P=0.454$), but there was a difference in EvoLang 11 (difference in $means=0.17$, $t=4.4$, $P < 0.0001$). In other words, female first-authored papers ranked higher in the conference with DBR.

The results suggest that DBR helps reduce a bias against female authors.

Regarding student status, in general, student papers were rated as better than non-student papers, which was less prominent in EvoLang 11. This might be explained by authors being more lenient towards student papers (or, conversely, more critical of minor problems by established authors), and this effect is then minimized under DBR. That is, there was no evidence of the Matthew effect in the overall data.

\bigskip
\subsection{Citation gap: with and without identifiable gender} \label{PC}

\medskip
\section*{The Gender Citation Gap in International Relations}
\medskip

In 2013, Maliniak et al. \cite{maliniak2013} did a study on how much citation and publication patterns differ between men and women in the international relations literature. 

\begin{figure}
\centering
\includegraphics[width=.48\textwidth,clip]{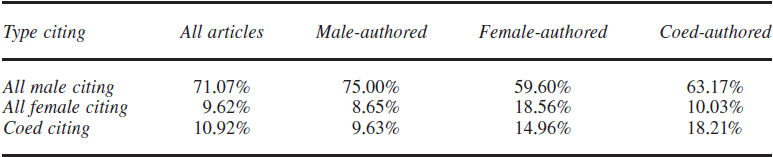}
\caption{Dyadic citations by gender, percentages represent the mean for all articles of each type.} \label{IR}
\end{figure}

Using data from the Teaching, Research, and International Policy project on peer-reviewed publications between 1980 and 2006, it is shown that women are systematically cited less than men after controlling for a wide range of factors, as suggested by the results in Table. \ref{IR}. 
A research article written by a woman and published in any of the top journals will still receive significantly fewer citations than if a man had written that same article.

Articles authored by women are systematically less central than articles authored by men, all else equal. This is likely because 1. women tend to cite themselves less than men, and 2. men (who make up a disproportionate share of IR scholars) tend to cite men more than women.

The findings in this paper offer robust evidence for a gender gap in citation counts in IR. This is a cause for concern.
If women in IR are systematically cited less than men in ways that do not appear to be associated with observable differences in their scholarship, and if citation counts continue to be used as a key measure of research impact, then women will be disadvantaged in tenure, promotion,
and salary decisions.

Some of the conclusions in this paper follow.

First, citation counts are not a fair and objective measure of the quality and impact of a scholar.
We now know that women will have lower citation counts than their male colleagues, all else equal.
Moreover, the bias stems not from a difference in quality, topic, or choice of research strategy but from certain underlying behaviors (fewer self-citations by women and more within gender citations).

Second, self-promotion strongly affects citation counts, and women are less likely to promote themselves.

Third, scholars tend to cite work by scholars of the same gender. For a field heavily dominated by men, as is the case in IR (and, regarding the current paper, in STEM), this pattern will lead to significantly fewer citations for women. It also means that the gap is not likely to disappear until a more equal number of male and female researchers exists.

Finally, networks matter. Producing high-quality work is insufficient for research to gain the attention of the widest number of scholars or have the greatest impact. If networks tend to bifurcate along gender lines, then any field that is disproportionately male will also disproportionately favor their work.

\medskip
\section*{Gender differences and bias in open source: pull request acceptance of women versus men}
\medskip

We evaluate a second thorough study by Terrel et al. in 2017 \cite{Terrel2017} on gender differences and bias in open source, which analyses pull request acceptance of women versus men on GitHub. 

Here, results show that women's contributions tend to be accepted more often than men's. However, for contributors who are outsiders to a project and whose gender is identifiable, men's acceptance rates are higher. Results suggest that although women on GitHub may be more competent overall, bias against them exists nonetheless.

\begin{figure}
\centering
\includegraphics[width=.48\textwidth,clip]{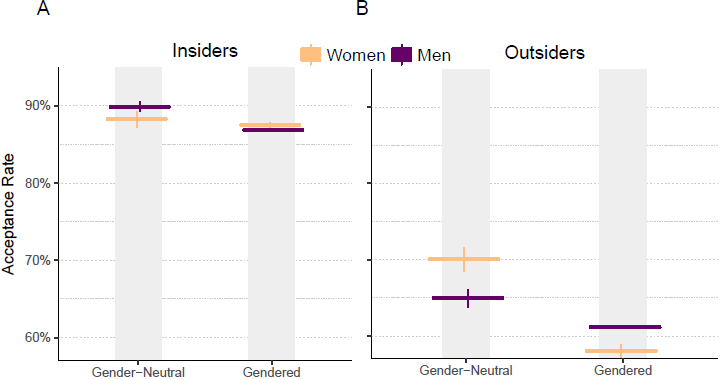}
\caption{Pull request acceptance rate by gender and perceived gender, with 95\% Clopper-Pearson confidence intervals, for insiders (A) and outsiders (B).} \label{github}
\end{figure}

In this paper, the author discovers that women's contributions are, in percentage, more frequently accepted than men's contributions. The author then evaluates several hypotheses that may justify the difference; some examples are: 
\begin{itemize}
    \item \textit{Do women’s pull request acceptance rates start low and increase over time?} No, between 1 and 64 pull
requests, women's higher acceptance rate remains.
    \item \textit{Are women focusing their efforts on fewer projects?} Yes, but not relevant because, when contributing to between 1 and 5 projects, women have a higher acceptance rate as they contribute to more projects.
    \item \textit{Are women making pull requests that are more needed?} Contrary to the hypothesis, women are slightly less likely to submit a pull request that mentions an issue, suggesting that women's pull requests are less likely to fulfill a documented need. Regardless, the result suggests that women's increased success rate is not explained by making more specifically needed pull requests.
    \item \textit{Are women making smaller changes?} No, for three of four measures of size, women's pull requests are significantly larger than men's.
    \item \textit{Are women’s pull requests more successful when contributing code?} No, women's acceptance rates are higher than men's for almost every programming language. The one exception is .m, which indicates Objective-C and Matlab, for which the difference is not statistically significant.
    \item \textit{Are acceptance rates different if we control for covariates?} They are not different regarding the type of project, the type of language (in which Ruby, Python, and $C++$ are significantly higher for women, and PHP is significantly higher for men), the pull request index, and the number of pull requests per person.
    \item \textit{Is a woman’s pull request accepted more often because she appears to be a woman?} We evaluate this hypothesis by comparing the pull request acceptance rate of developers who have gender-neutral GitHub profiles and those who have gendered GitHub profiles, and the results follow in Fig. \ref{github}.
\end{itemize}

Analysing Fig. \ref{github}, for insiders, we observe little evidence of bias when we compare women with gender-neutral profiles and women with gendered profiles since both have similar acceptance rates.
This can be explained by the fact that insiders likely know each other to some degree since they are all authorized to make changes to the project and thus may be aware of each others' gender.

For outsiders, we see evidence for gender bias: women's acceptance rates drop by 12.0\% when their gender is identifiable, compared to when it is not.
There is a smaller 3.8\% drop for men. Women have a higher acceptance rate of pull requests overall (as we reported earlier), but when they are outsiders and their gender is identifiable, they have a lower acceptance rate than men.



In the context of existing theories of gender in the workplace, plausible explanations include the presence of gender bias in open source, survivorship and self-selection bias, and women being held to higher performance standards.

\bigskip
\subsection{Biased exam grading: before and after anonymous evaluation} \label{EG}

\medskip
\section*{The Genius is a Male: Stereotypes and Same-
Sex Bias in Exam Grading in Economics at Stockholm University}
\medskip

In 2020, Jansson et al. \cite{Jansson2018} wrote an article that depicts gender bias in exam grading in Economics at
Stockholm University, in Sweden. This paper found evidence of same-sex bias before anonymous exams were introduced. Also, once anonymous grading was in place, the
effect disappeared. More specifically, when separating the effects by grader's sex, both groups of graders favor male
students, although male graders favor male students to a larger extent than female graders. They add that there was no evidence of compositional changes across the pre-and post-anonymous grading regimes.

The data collected was from the macroeconomics exam, consisting of essay and multiple choice questions, of the introductory course at Stockholm University from the spring of 2008 to the fall of 2014. 

First, the graders were randomly assigned to the 7 essay questions by ballot, and so, the characteristics of the students were balanced across the gender of the graders both in the pre-and post-anonymity sample.

Second, there is no statistical evidence of female students becoming smarter or older than male students in the post-reform compared to the pre-reform period.

\begin{figure}
\centering
\includegraphics[width=.48\textwidth,clip]{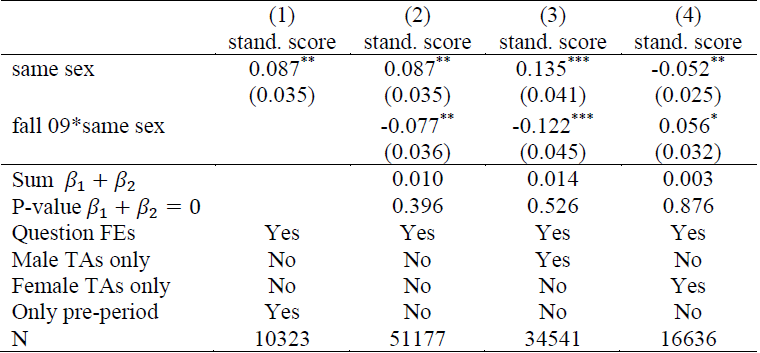}
\caption{Results for same-sex bias.} \label{economics}
\end{figure}

In Table \ref{economics}, Column 1 shows that being corrected by a grader of the same sex increased the exam score by 0.087 standard deviations when the exams were not anonymously graded. Reassuringly, this same-sex bias disappeared once anonymous exams were introduced, as the interaction is approximately the same size as the pre-reform effect (column 2).

Columns 3 and 4 separate the sample and analyze male and female graders separately. The male graders scored male students 0.14 st.d. higher than female students. Once anonymous exams were used, the effect was again close to zero. 
However, female graders scored female students significantly worse than male students (0.052 st.d.), and the effect once again was close to zero when exams were anonymous. 

The conclusion is the existence of a grading bias mainly against female students. The authors consider that this is consistent with the theories of how sex stereotypes (genius is male) affect judgment.

Moreover, the findings imply that equal sex representation among university teachers would not necessarily provide unbiased grading at a group level.
Furthermore, these results directly prove the effectiveness of anonymous evaluation.

Indeed, even though the results may seem surprising and we might have never thought about bias in exam grading, we should all consider implementing anonymous evaluation in our course's exams.

\section{Discussion: Counter arguments}\label{discussion}

\subsection{Is Double-blind review always a miraculous solution?}

Some studies found no difference in gender balance as a result of DBR \cite{critic2008, counter2009}. 
Webb et al. \cite{doesit2008} and Engqvist and Frommen \cite{counter2008} argue that the increase in ratings of female-authored papers is partly caused simply by an increasing number of females in the pool of submitters or a general reduction in bias over time, rather than an effect of review type.

In 2018, Cox and Montgomerie \cite{prosandcons2018} wrote a very interesting paper that intends to give continuity to the work of Budden et al. in 2008 \cite{BUDDEN20084} and challenge it. 
They tested whether double-blind reviews have influenced the recent incidence of female authorships in the journal \textit{BE}.

They analyzed the percentage of female authorships in each issue of each journal. 
The chosen period was 2010 to 2018 to provide a current estimate of gender biases and recent trends, specifically to update the information in Budden et al. \cite{BUDDEN20084} which covers papers published in \textit{BE} and other journals from 1997 to 2005.

\subsection{Double blind review is an obvious solution, so why do so few journals outside of medicine, psychology, and economics \cite{peer2005} practice this policy?}

One concern is that it might be possible to infer the author's identity from information such as the study system or cited
work \cite{moller2001}. However, evidence from other disciplines has shown that when encouraged to guess the identity of the
author or institution, reviewers were correct on only 25–42\% of occasions \cite{blind1991, blind2002}. 

Other concerns include the decreased ability to detect publication of the same data across multiple papers \cite{blind1995}, lost potential for increased feedback to more junior authors \cite{future2005}, and concerns regarding increased workload associated with double-blind review \cite{blind1995, future2005, notblind2006}.
Also, it involves an increased administrative load.


\section{CONCLUSIONS}\label{conclusion}

The first hypothesis regarding the existence of bias against women in scholarly contributions has been illustrated in many studies with gender-segregated publications data. 
Some of these studies conclude that using their available data set and testing for multiple scenarios, this hypothesis is verified to different levels in each study.
Some studies deem that, at least for some journals, the results for the available data are inconclusive or even suggest this bias is undetected or unapparent.

These studies span from STEM to academia to economics, linguistics, and others. Most of them found a discrepancy in success when the gender of the first author is identifiable versus when it's clearly a woman's contribution. Therefore, women appear to be more likely to get their papers rejected and not cited as often as men.  

We also proposed two simple modifications to tackle gender bias in STEM that can be applied to (but not only) IEEE conferences and journals. These modifications were supported by two hypotheses regarding paper acceptance and paper citation. After proving these hypotheses, we propose their implementation in IEEE conferences and journals as a necessary measure towards the improvement of the gender gap in academia and STEM.

The two proposed measures are the implementation of a double-blind review and one  letter to identify the authors' first names, followed by their family names. The first one has been applied in many conferences and journals, and studies on its efficacy have been made. The second appears to be applied in an informal and mixed manner, depending on the publishing institutions, research groups, or even personal preferences. We propose to implement this change in a mandatory fashion, standardizing the format of author names in publications.

We also propose other modifications regarding gender bias in STEM and academia, such as anonymous exam grading, and encourage further reforms supported by current research on this topic with gender-segregated data.


\bibliographystyle{IEEEtran}
\bibliography{references}

\end{document}